\documentclass{elsarta}
\usepackage{graphicx}
\begin{document}
\hspace*{3.5 in}CUQM-112\\
\hspace*{3.5 in}math-ph/0512079\\
\hspace*{3.5 in}December 2005
\vspace*{0.4 in}
\begin{frontmatter}
\title{Exact solutions for semirelativistic problems with non-local potentials}
\author{Richard L. Hall}
\ead{rhall@mathstat.concordia.ca}
\address{Department of Mathematics and Statistics, Concordia University,
1455 de Maisonneuve Boulevard West, Montr\'eal,
Qu\'ebec, Canada H3G 1M8}
\begin{abstract}
 It is shown that exact solutions may be found for the energy eigenvalue problem generated by the class of semirelativistic Hamiltonians of the form $H = \sqrt{m^2+p^2} + \hat{V},$ where $\hat{V}$ is a non-local potential with a separable kernel of the form ${\mathcal V}(r,r') = - \sum_{i=1}^n v_i f_i(r)g_i(r').$ Explicit examples in one and three dimensions are discussed, including the Yamaguchi and Gauss potentials. The results are used to obtain lower bounds for the energy of the corresponding $N$-boson problem, with upper bounds provided by the use of a Gaussian trial function. 
\end{abstract}

\end{frontmatter}
\section{Introduction}
We study semirelativistic problems in which the Hamiltonian $H$ has the relativistically
correct expression $K(p^2) = \sqrt{m^2 + p^2},$ $p \equiv |{\bf p}|,$ for
the energy of a free particle of mass $m$ and momentum ${\bf p},$ and an added
static interaction potential $\hat{V}.$  The Hamiltonian is therefore given by
$$H = \sqrt{m^2 + p^2} + \hat{V}.\eqno{(1.1)}$$
The eigenvalue equation $H\psi = E\psi$ is usually called the spinless Salpeter equation \cite{bse} \cite{se}. For many potentials, this Hamiltonian can be shown \cite{lieb} to be bounded below and essentially self-adjoint, and its spectrum can be defined variationally. From the point of view of solvability, these features represent  significant technical advantages over the more-complete Bethe-Salpeter formulation. There is, however, one remaining difficulty, namely the non-locality of the kinetic-energy operator.

The `usual' multiplicative potential operator of elementary quantum mechanics is generated by a special kernel of the form ${\mathcal V}(x,x')=V(x)\delta(x,x').$  Thus we have
$$(\hat{V}\psi)(x) = \int\limits_{-\infty}^{\infty}V(x)\delta(x,x')\psi(x')dx' = V(x)\psi(x),\eqno{(1.2)}$$
and this special form makes $\hat{V}$ a local `multiplicative' operator.  Since (with $\hbar = 1$) the Schr\"odinger kinetic-energy operator $p^2/(2m) = -\partial_x^2/(2m)$ is also local, the non-relativistic Hamiltonian is a local operator.  By contrast, the kinetic-energy operator $\hat{K} = \sqrt{m^2 + p^2}$ in the semirelativistic problem is non-local and this is the source of many of the difficulties encountered with the corresponding eigenvalue problem. The action of $\hat{K}$ is defined \cite{lieb} in terms of the Fourier transform ${\mathcal F}(\psi) = \tilde{\psi}.$ Thus, in one dimension, we have explicitly: 
$${\mathcal F}(\hat{K}\psi)(k) =  \sqrt{m^2 + k^2}~\tilde{\psi}(k),\eqno{(1.3)}$$
where
$$\tilde{\psi}(k) = \langle\psi|k\rangle = \int\limits_{-\infty}^{\infty}\langle\psi|x\rangle dx \langle x|k\rangle = {1\over{\sqrt{2\pi}}}\int\limits_{-\infty}^{\infty} e^{-ikx}\psi(x)dx.\eqno{(1.4)}$$
The main purpose of the present article is to show that with separable potentials, the non-locality of {\it both} terms in the Hamiltonian allows us to solve the eigenproblem exactly, up to a definite integral. This result, in turn, allows us to find a lower bound to the energy of the corresponding $N$-boson problem in which the particles interact pairwise.  The class of potential kernels we shall consider may be written (for the one-body problem in one dimension) 
$${\mathcal V}(x, x') = -v\sum\limits_{i = 1}^{n}f_i(x)g_i(x').\eqno{(1.5)}$$
Such potentials have been studied as models for a variety of physical problems \cite{yamag,muk,levay,balan,kidun}. Our main general results for a single particle in one and three dimensions are proved in Section~2.  In Section~3 we look at some exponential examples in one dimension and in Section~4 we solve the eigenproblem for the three-dimensional Yamaguchi \cite{yamag} and Gauss potentials. In Section~5 we apply the results to study a system of $N$ identical bosons interacting pairwise in three dimensions via a non-local Gauss potential: the one-particle exact solutions provide an energy lower bound, to which we adjoin a variational upper bound derived with the aid of a scale-optimized Gaussian trial function.
\section{Exact solutions}
For definiteness, we first solve the problem with one separable potential term in one spatial dimension. Thus we suppose that the kernel of the potential operator $\hat{V}$ has the form
$${\mathcal V}(x, x') = -v f(x) g(x'),\eqno{(2.1)}$$
where $v$ is a postive coupling parameter. The eigen equation for the semirelativistic one-body problem becomes
$$\sqrt{m^2 + p^2}~\psi(x) - \int\limits_{-\infty}^{\infty}v f(x) g(x')\psi(x')dx' = E\psi(x).\eqno{(2.2)}$$
If we represent the Fourier transforms by ${\mathcal F}(\psi) = \tilde{\psi},$${\mathcal F}(f) = \tilde{f},$ and ${\mathcal F}(g) = \tilde{g},$ then Eq.(2.2) becomes
$$ \sqrt{m^2 + k^2}~\tilde{\psi}(k) - v c\tilde{f}(k) = E\tilde{\psi}(k),\eqno{(2.3)}$$
where the constant $c$ is given by
$$c = \int\limits_{-\infty}^{\infty}\tilde{g}(k')\tilde{\psi}(k')dk'.$$
Thus $\tilde{\psi}$ is given by
$$\tilde{\psi}(k) = {{cv\tilde{f}(k)}\over{\sqrt{m^2 + k^2} - E}}.\eqno{(2.4)}$$
If we now multiply both sides of (2.4) by $\tilde{g}(k)$ and integrate, we find the following formula relating the reciprocal coupling to the energy $E$:
$${1\over{v}} = \int\limits_{-\infty}^{\infty}{{\tilde{f}(k)\tilde{g}(k)dk}\over{\sqrt{m^2+k^2} - E}}.\eqno{(2.5)}$$
Equations (2.4) and (2.5) show that if there is a solution for given $f(x)$ and $g(x),$ then this solution is unique (up to a phase). The  corresponding energy eigenvalue is now determined by (2.5) since $v$ is a monotone function of $E$. It is clear that $\sqrt{m^2 + k^2} \geq m.$  ~If we write $E = E(m) = m + e(m),$ then, for bound states, $e = E - m < 0.$ Consequently we have $\sqrt{m^2 + k^2} - E > 0.$ That is to say, there are no mathematical singularities arising from this factor in the various integrands.  In the large-$m$ limit, the problem approaches the corresponding non-relativistic case since $\sqrt{m^2 + k^2}-m \sim k^2/(2m);$ moreover, this approach is from below since  $\sqrt{m^2 + k^2}-m < k^2/(2m).$ In the examples we shall consider, the function $E(m) - m$ approaches the non-relativistic $m$-dependence as the mass increases.  Another interesting special case for the semirelativistic problem is the ultra-relativistic limit $m\rightarrow 0.$ This, of course, has no natural non-relativistic counterpart.

When there are more than one term in the separable potential we have
$${\mathcal V}(x, x') = -\sum\limits_{i = 1}^{n}v_i f_i(x) g_i(x'),\eqno{(2.6)}$$
We now define the constants $\{c_i\}_{i=1}^n$ by
$$c_i = \int\limits_{-\infty}^{\infty}\tilde{g}_i(k')\tilde{\psi}(k')dk',\eqno{(2.7)}$$
and the formula (2.4) for the wave function in this more-general case becomes
$$\tilde{\psi}(k) = {{\sum\limits_{i = 1}^{n}v_i\tilde{f}_i(k)c_i}\over{\sqrt{m^2+k^2} - E}}.\eqno{(2.8)}$$
The relation between the coupling parameters and the eigenvalue is now expressed by the condition that the linear equations for the constants $\{c_i\}_{i=1}^n$ are non trivial.  If we define the matrix elements of the $n\times n$ matrix $J$ by
$$J_{ji} = v_i\int\limits_{-\infty}^{\infty}{{\tilde{g}_j(k)\tilde{f}_i(k)dk}\over{\sqrt{m^2 + k^2} - E}}, \eqno{(2.9)}$$
then the more-general eigenvalue formula, corresponding to (2.5), may be written
$$\det(I-J)= 0.\eqno{(2.10)}$$

There are similar results in three spatial dimensions.  We consider one-term central potentials of the form
$${\mathcal V}({\bf r}, {\bf r'}) = -v f(r) g(r'),\eqno{(2.11)}$$
where $r = |{\bf r}|.$  In this case the Fourier transform ${\mathcal F}(f)$ of $f$, for example, takes the form
$$\tilde{f}(k) = {1\over{(2\pi)^{3\over 2}}}\int e^{-i{\bf k}\cdot{\bf r}}f(r)d^3{\bf r} = \left({2\over{\pi}}\right)^{1\over 2}{1\over k}\int\limits_0^{\infty}\sin(k r) r f(r) dr\eqno{(2.12)}$$
in which $k = |{\bf k}|.$ Similar reasoning to that of the one-dimensional case then yields the solution formulae
$$\tilde{\psi}(k) = {{cv\tilde{f}(k)}\over{\sqrt{m^2 + k^2} - E}}\eqno{(2.13)}$$
and
$${1\over{v}} = 4\pi \int\limits_{-\infty}^{\infty}{{\tilde{f}(k)\tilde{g}(k)k^2 dk}\over{\sqrt{m^2+k^2} - E}}.\eqno{(2.14)}$$
These results can also easily be extended to potential kernels with a sum of separable terms.
\section{Problems in one dimension}
We now consider some examples.  Since the general solution is given in Section~2, the purpose of the examples is to demonstrate that exact solutions are indeed feasible.  We solve the first problem in some detail and then present summary solutions and results for a selection of other problems.   
\subsection{The one-term exponential potential}
We consider the potential 
$${\mathcal V}(x,x') = -v f(x)f(x') = -v e^{-|x|/a} e^{-|x'|/a},\quad v,\ a\ >0.\eqno{(3.1)}$$
The potential factors in momentum space are therefore given by
$$\tilde{f}(k) = \sqrt{{2\over {\pi}}}\int\limits_0^{\infty}\cos(kx)f(x)dx = 
\sqrt{{2\over {\pi}}}\left({a\over{1+a^2 k^2}}\right),\eqno{(3.2)}$$
and the formulae (2.4) and (2.5) for the momentum-space wave function and the corresponding eigenvalue become
$$\tilde{\psi}(k) = {{cav}\over{\left(1+a^2k^2\right)\left(\sqrt{m^2+k^2}- E\right)}}\eqno{(3.4)}$$
and
$${1\over{v}} = \int\limits_{-\infty}^{\infty}{{\tilde{f}^{2}(k)dk}\over{\sqrt{m^2+k^2} + |E|}} = {{4a^2}\over{\pi}}\int\limits_0^{\infty}
{1\over{\left(1+a^2k^2\right)^2 \left(\sqrt{m^2+k^2} - E\right)}}.\eqno{(3.5)}$$
This equation may be inverted to give $E$ for each choice of the parameter set $\{a,m,v\}.$  In Fig.(1) we exhibit the $m$ dependence of $E-m$ for $a = 1$ and $v = \{1, 2, 3\}.$  In the Schr\"odinger limit, $m\rightarrow \infty$, we find
$$e(m) = E(m) - m = -\frac{v}{a}.\eqno{(3.6)}$$
Meanwhile for the ultrarelativistic case $m = 0$ we have 
$${1\over{v}} = -\frac{a^2\left(2 + 2a^2e^2+3ae\pi+a^3e^3\pi+4\ln(-ae)\right)}{(1+a^2e^2)^2\pi},\quad e < 0.\eqno{(3.7)}$$
The graphs shown in  Fig.(1) are consistent with these relations.  We see that this semirelativistic problem is indeed exactly soluble.
\begin{figure}[htbp]
\centering
\includegraphics[width=12 cm]{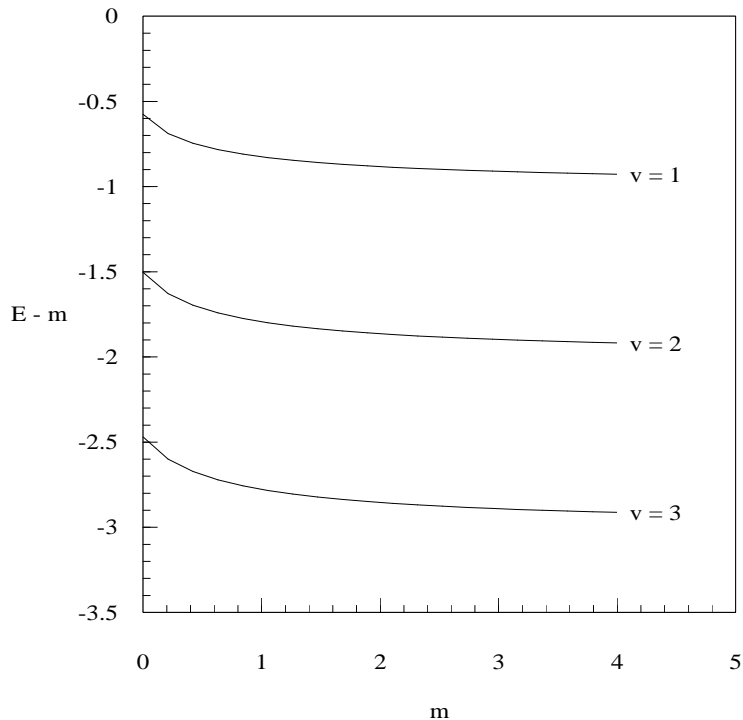}
\caption{Plots of exact semirelativistic energies $E(m) - m$ for the nonlocal exponential potential 
${\mathcal V}(x,x') = -v e^{-|x|-|x'|}$ for three values of the coupling $v.$}
\label{Fig. (1)}
\end{figure}
\subsection{A two-term exponential potential}
We consider now the case 
$${\mathcal V}(x,x') = -v_a e^{-(|x|+|x'|)/a}-v_b e^{-(|x|+|x'|)/b},\quad v_a,\ v_b,\ a,\ b\ > 0.\eqno{(3.8)}$$ 
In particular, if we choose the explicit values $a = 1, b = 2, v_a = v_b = 1,$ the secular equation (2.10) becomes
$$(1-J_{11})(1-J_{22}) - J_{12}^2 = 0,\eqno{(3.9)}$$
where the integrals are given by
$$J_{11} = {4\over{\pi}}\int\limits_0^{\infty}{dk\over{(1+k^2)^2(\sqrt{m^2+k^2}-E)}},\eqno{(3.10a)}$$
$$J_{22} = {16\over{\pi}}\int\limits_0^{\infty}{dk\over{(1+4k^2)^2(\sqrt{m^2+k^2}- E)}},\eqno{(3.10b)}$$
and
$$J_{12} = {8\over{\pi}}\int\limits_0^{\infty}{dk\over{(1+k^2)(1+4k^2)(\sqrt{m^2+k^2}- E)}}.\eqno{(3.10c)}$$
Thus for $m = \{0,~ 0.5,~ 1\}$ we find respectively from (3.9) that $E = \{-1.14462,~ -0.814543,~ -0.36131\}.$  In Fig.(2) we exhibit a graph showing $E -m$ as a function of $m$ for this problem. 
\begin{figure}[htbp]
\centering
\includegraphics[width=12 cm]{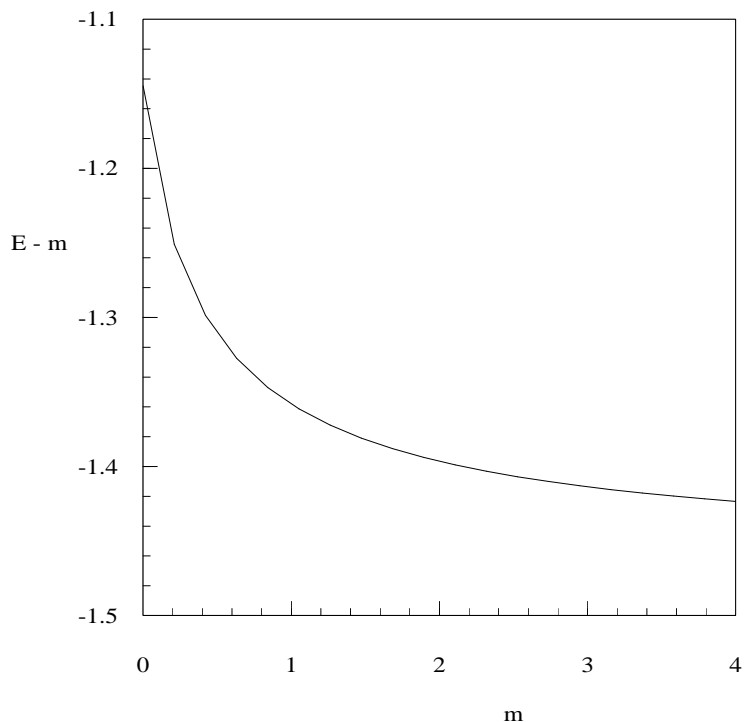}
\caption{Plots of exact semirelativistic energies $E(m) - m$ for the nonlocal two-term exponential potential 
${\mathcal V}(x,x') = -e^{-(|x|+|x'|)}-e^{-\frac{1}{2}(|x|+|x'|)}.$ }
\label{Fig. (2)}
\end{figure}
\section{Problems in three dimensions}
The Yamaguchi potential \cite{yamag} has a potential kernel given by
$${\mathcal V}({\bf r},{\bf r'}) = -v \left(\frac{e^{-\beta r}}{r}\right)\left(\frac{e^{-\beta r'}}{r'}\right), \quad v,~ \beta >0.\eqno{(4.1)}$$
Because of the volume measure $r^2 dr$ in three dimensions cancels the singularities in the Yukawa-type factors, this problem is very similar to the exponential potential in one dimension.  The wave function and eigenvalue formula are found from (2.13) and (2.14) to be respectively
$$\tilde{\psi}(k) = \frac{c}{\left(k^2 + \beta^2\right) \left(\sqrt{m^2+k^2} - E\right)}\eqno{(4.2)}$$
and
$$\frac{1}{v} = 8 \int\limits_0^{\infty}\frac{k^2dk}{\left(k^2 + \beta^2\right)^2 \left(\sqrt{m^2+k^2} - E\right)}.\eqno{(4.3)}$$
Thus the energy $E$ may be found from (4.3) as a function of the positive parameters $\{m, \beta, v\}.$

Similarly, for the Gauss potential we have the kernel
$${\mathcal V}({\bf r},{\bf r'}) = -v e^{-\half\beta (r^2+ r'^2)}\quad v,~ \beta >0.\eqno{(4.4)}$$
The corresponding wave function and eigenvalue formula in this case are given by
$$\tilde{\psi}(k) = \frac{c e^{-\half k^2/\beta}}{\sqrt{m^2+k^2} - E}\eqno{(4.5)}$$
and
$$\frac{1}{v} = \frac{4 \pi}{\beta^3} \int\limits_0^{\infty}\frac{e^{-k^2/\beta}k^2dk}{\sqrt{m^2+k^2} - E}.\eqno{(4.6)}$$
\section{The semirelativistic $N$-boson problem}
In this section we consider a system of $N$ identical bosons interacting pairwise in three spatial dimensions.  The Hamiltonian for the system may be written
$$H= \sum_{i=1}^N(m^2+p_i^2)^{\half}+\sum_{j>i=1}^N \hat{V}_{ij},\eqno{(5.1)}$$
where, for a single particle, the action of the Gauss potential is given by 
$$\hat{V}\psi({\bf r}) = -v\int e^{-\frac{\beta}{2}(r^2+r'^2)}\psi({\bf r'})d^3{\bf r'},\quad v > 0.\eqno{(5.2)}$$
We shall consider two distinct approaches.  First we obtain a lower bound to the lowest $N$-body energy $E$ with the aid of a scaled one-body problem and secondly we find an upper bound with the aid of an $N$-body Gaussian trial wave function.
\subsection{The lower bound} 
If we suppose that $\Psi$ is the exact (unknown) $N$-boson wave function, then boson symmetry implies that $E = (\Psi, H\Psi)= (\Psi,h\Psi),$ where $h$ is a two-body Hamiltonian given by
$$h = \frac{N}{2}\left[(m^2+{\bf p}_1^2)^{\half}+ (m^2+{\bf p}_2^2)^{\half} + (N-1)\hat{V}_{12}\right]\eqno{(5.3)}.$$
If new coordinates for the two-body problem are ${\bf r} = {\bf r}_1-{\bf r}_2$ and $\rho ={\bf r}_1+{\bf r}_2,$ then the individual momenta are given by $\pi\pm{\bf p},$ and, by using the lemma of Ref. \cite{halllemma} to `remove' the operator $\pi$ from within expectation values, the two-body operator $h$ may be replaced by ${\mathcal H},$ where
$${\mathcal H} = N\left[(m^2+p^2)^{\half} + \half(N-1)\hat{V}\right],\eqno{(5.4)}$$  
Thus we conclude that
$$E = (\Psi, H\Psi)= (\Psi,{\mathcal H}\Psi) \geq {\mathcal E} = E_{L},\eqno{(5.5)}$$
where ${\mathcal E}$ is the bottom of the spectrum of the one-body operator ${\mathcal H}.$ By comparing (5.4) with (4.6) we see that
 $$\frac{1}{(N-1)v} = \frac{2 \pi}{\beta^3} \int\limits_0^{\infty}\frac{e^{-k^2/\beta}k^2dk}{\sqrt{m^2+k^2} - E_{L}/N}.\eqno{(5.5a)}$$
Thus, for each choice of the parameters $m$ and $\beta$, (5.5a) implies that $E_{L}/N$ is a function of $v(N-1).$ In the special case $m = \beta = 1,$ we write this function as $f_L$ so that we have
$$E_L/N = f_L(v(N-1)).\eqno{(5.5b)}$$
We note the special critical coupling $u_c$ defined by $f(u_c) = 0$ is given by $v(N-1)|_c = 0.527485.$
\subsection{The upper bound}
\begin{figure}[htbp]
\centering
\includegraphics[width=12 cm]{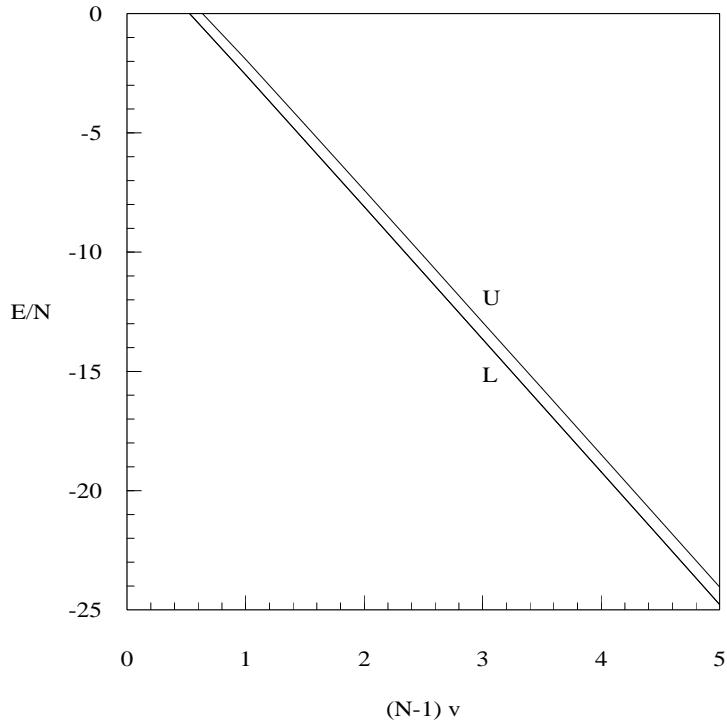}
\caption{Upper and lower bounds for $E/N$ against $(N-1) v$ for $N$ identical bosons interacting via the Gauss pair potential ${\mathcal V}(r,r') = -e^{-\frac{\beta}{2}(r^2 + r'^2)}$ . The lower curve (L) is for $N = 2$ and the upper curve (U) is for $N = \infty;$ the curves for $2<N<\infty$ lie between these two.}
\label{Fig. (3)}
\end{figure}
For a variational upper bound we adopt explicit relative coordinates. Jacobi coordinates may be defined with the aid of an orthogonal 
matrix $B$ relating the column vectors of the new $[\rho_i]$ and old 
$[{\bf r}_i]$ coordinates given by $[\rho_i]=B[{\bf r}_i].$ The 
first row of $B$ defines a center-of-mass variable $\rho_1$ with every entry 
$1/\sqrt{N},$ the second row defines a pair distance $\rho_2 
=({\bf r}_1-{\bf r}_2)/\sqrt{2},$ and the $k\!$th row, $k\ge 2,$ has the first 
$k-1$ entries $B_{ki}=1/\sqrt{k(k-1)},$ the $k\!$th entry 
$B_{kk}=-\sqrt{(k-1)/k},$ and the remaining entries zero. We define the 
corresponding momentum variables by $[\pi_i]=(B^{-1})^{\rm 
t}[{\bf p}_i]=B[{\bf p}_i].$ The trial wave function we use is given by
$$\Phi(\rho_2,\rho_3,\dots,\rho_N)= 
C e^{-{{\alpha}\over{2}}\sum_{i=2}^N\rho_i^2} = C\prod\limits_{i = 2}^N\phi(\rho_i)\eqno{(5.6)}$$
where $\alpha > 0$ and
$C$ is a normalization constant.  This function is symmetric in the individual position coordinates $\{{\bf r}_i\}_{i=1}^N$ and also in the $N-1$ relative coordinates $\{\rho_i\}_{i = 2}^N;$ meanwhile it has the unique factoring property shown.  These facts enable us \cite{halllemma} to express the expectation of the full Hamiltonian $H$ in the form
$$E \leq (\Psi, H\Psi) = N(\phi,\ ((m^2 + 2\lambda p^2)^{\half} + (N-1) \hat{V})\phi),\eqno{(5.7)}$$
where $\lambda = (N-1)/N,$ the potential operator $\hat{V}$ has the Gauss kernel (4.4), and $\alpha$ is to be used as a variational parameter.  We therefore obtain the following expression for the upper bound $E_U$ in the special case $m = \beta = 1$
$$\frac{E}{N} \leq \frac{E_U}{N} = \left(\frac{2}{\pi}\right)^{\half}\min_{s > 0}\left[\frac{g(s^2)}{s} - 8 v \pi^2\frac{(2\lambda s^2)^{\frac{3}{2}}}{(1+4\lambda s^2)^3}\right],\eqno{(5.8a)}$$
where the monotone function $g$ is given by
$$g(x) = \int_{-\infty}^{\infty}e^{-t^2}\left[2x+t^2\right]^{\half}t^2 dt = x e^{x} K_1(x).$$
In this last expression, $K_{\nu}(x)$ is a modified Bessel function of the second kind \cite{abram}. The result of the minimization in (5.8a) yields $E_U/N$ as a function $f_U$ of $(N-1) v$ and $\lambda,$ where $\half\leq\lambda\leq 1.$ We have
$$E_U/N = f_U(v(N-1),\lambda).\eqno{(5.8b)}$$
Thus we obtain a different upper-bound curve for each $\lambda = (N-1)/N.$ These curves do not intersect.  In Fig.(3) we exhibit the lower curve $f_L(v(N-1)),$ valid for all $N$, the upper curve  $f_U(v(N-1),\half),$ for $N = 2,$ and the upper curve $f_U(v(N-1),1),$ for $N = \infty.$  For the case $N=2$ the general lower (all $N\geq 2$) and particular upper bounds ($N=2$) are so close that they are indistinguishable on the graph: we have, for example, $f_L(1) = - 2.56844$ and $f_U(1,\half) = - 2.5651$ approximately.  Thus the scale-optimized Gaussian trial function is very effective for all $N,$ and particularly so for $N = 2.$  The apparent straightness of the energy curves can perhaps be understood by reasoning such as the following: for the lower bound (5.5a), the Gaussian in the integrand decays rapidly to zero, thus the mean-value theorem tells us, for a given $v$, that $(N-1)v = A - B(E_L/N);$ it remains, of course, to explain why $A$ and $B$ vary very slowly with $v.$However, with exact analytical results available (for both bounds), we do not have to look for more analytical approximations.
\section{Conclusion}
We have shown that exact solutions can be found to semirelativistic eigenvalue problems when the potential has a kernel that is a sum of separable terms.  This immediately extends, of course, to the wider class of $L^2$ kernels.   It may be possible to use such exact solutions to approximate the spectra generated by local potentials. The non-relativistic many-body problem with non-local potentials has already been studied \cite{halla} and the present paper extends these results to the corresponding semirelativistic case.  We have obtained tight bounds for the local semirelativistic $N$-body problem with local harmonic-oscillator potentials $V(r) = vr^2$, and somewhat weaker bounds for convex transformations $g(r^2)$ of the oscillator \cite{hallb}.  The work reported in the present paper will no doubt help us to extend these semirelativistic many-body results to wider classes of potentials.  It is very helpful when the lower bound itself, which is derived from a scaled one-body problem, can be found exactly. Improvements in the general lower bound await a treatment based on Jacobi relative coordinates; this has already been achieved in particular for the oscillator; the search for an improved general lower bound can now benefit from a non-oscillator test model for which there is also an accurate variational upper bound.
 \section*{Acknowledgement}
Partial financial support of this work under Grant No. GP3438 from the 
Natural Sciences and Engineering Research Council of Canada is gratefully acknowledged.

\end{document}